\documentclass[11pt,a4paper]{article}

\usepackage[utf8]{inputenc}
\usepackage{lmodern}

\usepackage{geometry}
\usepackage{graphicx}
\usepackage{caption}
\usepackage{subcaption}
\usepackage{booktabs}
\usepackage{tabularx}
\usepackage{array}
\usepackage{multirow}
\usepackage{longtable}
\usepackage{float}

\usepackage{amsmath, amssymb}

\usepackage{authblk}
\usepackage{setspace}
\usepackage{microtype}
\usepackage{enumitem}
\setlist[itemize]{noitemsep, topsep=0pt, parsep=0pt, partopsep=0pt, left=1.2em}

\usepackage[backend=biber, style=apa, citestyle=authoryear]{biblatex}
\usepackage{xcolor}
\usepackage[colorlinks=true, linkcolor=blue, citecolor=blue, urlcolor=blue]{hyperref}

\DeclareCiteCommand{\parencite}[\mkbibparens]
  {\usebibmacro{prenote}}
  {\textcolor{blue}{\usebibmacro{citeindex}%
   \usebibmacro{cite}}}
  {\multicitedelim}
  {\usebibmacro{postnote}}

\renewcommand{\arraystretch}{1.05}
\newcolumntype{Y}{>{\raggedright\arraybackslash}X}

\title{\textbf{Bibliometric-enhanced Systematic Literature Review of EEG in Education: Learning Concepts, Computational Methods, and Research Opportunities}}

\author[1]{Adi Wijaya}
\author[2]{Said Hasibuan}
\author[3,4]{Wiga Maulana Baihaqi}
\author[1,5]{Rizki Darmawan}
\author[6]{Rifkie Primartha}
\author[5]{Catur Supriyanto}

\affil[1]{Department of Health Information Management, Universitas Indonesia Maju, Jakarta, Indonesia}
\affil[2]{Department of Informatics, Institut Informatika dan Bisnis Darmajaya, Lampung, Indonesia}
\affil[3]{Department of Information Technology, Universitas Amikom Purwokerto, Purwokerto, Indonesia}
\affil[4]{Department of Electrical and Information Engineering, Universitas Gadjah Mada, Indonesia}
\affil[5]{Faculty of Computer Science, Universitas Dian Nuswantoro, Semarang, Indonesia}
\affil[6]{Department of Informatics, Universitas Sriwijaya, Palembang, Indonesia}

\date{}  

\begin{document}

\maketitle
\vspace{-1.2em} 
\noindent\textit{*Corresponding author: adiwjj@uima.ac.id}
\vspace{0.8em}   

\begin{abstract}
\noindent
Application of electroencephalography (EEG) in educational research has increased considerably, but an extensive integration of methodological frameworks, educational constructs, computational methodologies, and established research gaps remains an unexplored area. This study uses a Bibliometric-enhanced Systematic Literature Review (BenSLR) to provide an intensive and systematic overview of EEG applied in educational fields. Literature about the topic was extracted from Scopus and then systematically screened and analyzed, with keyword co-occurrence being evaluated using VOSviewer, with emerging trends represented through the Enhanced Strategic Diagram generated through BiblioPlot. The study found engagement, attention, and learning style to be salient constructs with machine learning and deep learning commonly utilized for modeling complex cognitive states. EEG signal processing, feature extraction, and assessment of cognitive and affective states continuously appeared across the systematically reviewed studies. Innovative interventions, including virtual reality and neurofeedback, illustrate how EEG is utilized in enabling adaptive and individualized educational experiences. Even with these developments, challenges continue in correlating neural markers with observable learning behavior, extending the measurement parameter beyond simple attention and working memory, and increasing the generalizability of predictive models. This study highlights the potential of the BenSLR methodology in systematically synthesizing literature, with qualitative and quantitative viewpoints delivered respectively. Furthermore, while in this study the BenSLR approach has been applied to EEG in education specifically, the methodology is flexible and transferable to other subjects or research area topics, assisting researchers in developing methodologies, theoretical frameworks, and evidence-based interventions.
\end{abstract}


\section{Introduction}
Electroencephalography (EEG) applied in education involves the utilization of EEG technologies to study and refine learning processes through the monitoring of electrical activity generated in the brain \parencite{chenApplicationElectroencephalographySensors2024}. The non-invasive instrument provides an efficient means of exploring cognition-related problems like engagement, attention, and cognitive load, thereby garnering educators’ significant insight into students' learning processes \parencite{apicellaEEGbasedMeasurementSystem2022}. The rising interest in EEG research in educational settings has been attributed to the ability of EEG to customize learning processes, finesse educational approaches, and garner immediate feedback from students and educators \parencite{liBiosensorTechnologyAdaptive2025a}. Following an appreciation of the neural substrates of learning through EEG, with better educational processes, has the potential to customize educational intervention based on the singular needs of individual students, with the net benefit of enhancing overall learning performances \parencite{grammerEffectsContextNeural2021a}. As educational frameworks continue to move towards more adaptive and individualized approaches, EEG has a pivotal role in determining the future of education \parencite{gkintoniChallengingCognitiveLoad2025}.\\

Recent studies have also confirmed the promise of EEG to study various aspects of learning. For instance, EEG has been used to classify learning style in terms of neural patterns to develop individualized learning experiences \parencite{yuvarajMachineLearningFramework2024a, zhangDesignImplementationEEGBased2021h,zhangTSMGDeepLearning2021e}. EEG has also been deployed to monitor students' attention while interacting with e-learning platforms to implement immediate changes to learning content and maintain appropriate levels of attention \parencite{elgammalNeuroTutorNeuralDecoding2025,niEEGBasedAttentionAnalysis2020a}. EEG-based indices of cognition, furthermore, combined with supervised learning of machines, have been shown to enhance educational outcomes through the provision of an instrument to measure and augment cognitive engagement \parencite{dursoEnhancingEducationalOutcomes2024b}. These experiments show that EEG can provide information and improve learning processes by providing clear and objective measurements of how a person thinks during the learning process.\\

The incorporation of EEG in educational research has increased exponentially, but so far, there is no total synthesis of dominant methodologies and approaches. The population is not just comprised of people but also of objects and other natural things. While students' psychological and cognitive responses are commonly measured with EEG, gaps in understanding their use across all branches of education still exist \parencite{orovasEEGEducationScoping2024a,tenorioBrainimagingTechniquesEducational2022}. To remedy these gaps, a systematic and thorough analysis of literature is required, not only to chart existing practices but also to ascertain future research avenues and generate standard methodologies \parencite{taranEducationalInitiativesImplementation2020}. For this, the Bibliometric-enhanced Systematic Literature Review (BenSLR) approach will be utilized. This approach integrates the classical SLR process with Bibliometric Analysis (BA) to narrow research questions and generate a systematic and comprehensive understanding of the topic.\\

This research endeavor is predicated on the recognition of existing deficiencies and the necessity for a systematic comprehension of educational studies utilizing EEG technology. Accordingly, the investigation intends to tackle three primary research questions. These questions have been designed to steer an exhaustive evaluation of the literature and will be elaborated upon and sharpened through the BenSLR methodology:\\

RQ 1: Which methodological trends prevail in educational research using EEGs? \par
RQ 2: What learning-related phenomena and educational constructs are most characteristically investigated with the use of EEG in the educational context? \par
RQ 3: What are the emerging trends, research gaps, and future avenues in educational research using EEG? \\

This research integrates heterogeneous research methodologies, prominent concepts from the educational realm, and corresponding processes in learning to present an expansive exploration of current uses of EEG-based research in educational settings. Further, it presents novel trends and subjects with more research needed. The research adopts a systematic approach to structure and reduce research questions and demonstrates how importing bibliometric methodologies with traditional systematic reviews becomes more effective in enhancing scholarly research specificity and richness. Practitioners and academics are expected to benefit from these findings to make better use of EEG, thereby fostering innovation in the learning process within educational settings \parencite{orovasEEGEducationScoping2024a}.
\section{Methods}
A systematic literature review (SLR) approach and bibliometric analysis were used in this study to systematically identify research questions. Research questions were effectively answered by finding, evaluating, and analyzing relevant previous articles. VOSviewer version 1.6.20 by \parencite{vanEck2010} was specifically adapted to perform keyword co-occurrence analysis, while BiblioPlot, developed by \parencite{wijayaBiblioPlotEnhancedData2025}, generated an Enhanced Strategic Diagram adapted from the work of \Parencite{shafin2022thematic} to help identify emerging trends, knowledge gaps, and future research directions. Qualitative and numerical information is provided with a comprehensive approach to produce a detailed description of the methods used in previous studies.\\
\subsection{Data Collection}
Data included in this study were extracted using a systematic search on July 18, 2025, on the Scopus database. We searched on titles, abstracts, and keywords (TITLE-ABS-KEY) using the following question: ("learning style" OR "learning behavior" OR "cognitive state" OR "learning attention" OR "learning engagement" OR "student engagement") AND (electroencephalogra* OR eeg OR brainwave OR brain-wave OR "brain wave" OR "brain rhythm") AND (student OR school OR college OR university OR e-learning OR "higher education" OR education OR academ*) AND NOT (review OR bibliometric OR editorial OR conference). This resulted in the selection of 235 articles published between 2004 and 2025 and was marked as potentially relevant records. The step-by-step selection process is detailed in Figure 1, derived from the PRISMA flow diagram \parencite{Page2021-qj}, but modified for this study to include extra components, including redundant records and the divergent pathway to bibliometric analysis and systematic literature review.\\

Of the initially collected articles, 12 non-technical articles and 5 non-English articles were deleted, leaving only 226 articles to be evaluated. However, after screening, only 110 were relevant to the research questions. These are spread across RQ1.1 (11), RQ1.2 (11), RQ2.1 (19), RQ2.2 (19), RQ2.3 (21), RQ3.1 (17), and RQ3.2 (12). However, 48 of the above records turned out to be duplicates, thus yielding a final inclusion of 62 unique articles. Detailed steps and results of each of the these screening phases are tabulated separately in the Paper Collection sub-section of the Results section, to ensure transparency and completeness of how the final dataset was derived. \\

The Scopus database is mainly used because of its practical benefits, especially when doing bibliometric analysis. Using Scopus helps keep the data organized consistently and provides a dependable set of data that can be used for more detailed checking and studies. Scopus also normally provides a high output of publications, beneficial for understanding the research landscape \parencite{zhu2020tale}. Its rich citation coverage was also reported in previous studies \parencite{pranckute2021wos}, and existing research supports its suitability for delineating hotspots of research and relevant trends within the field \parencite{zakaria2022blueberry}.\\

\begin{figure}[h]
    \centering
    \includegraphics[width=0.7\textwidth]{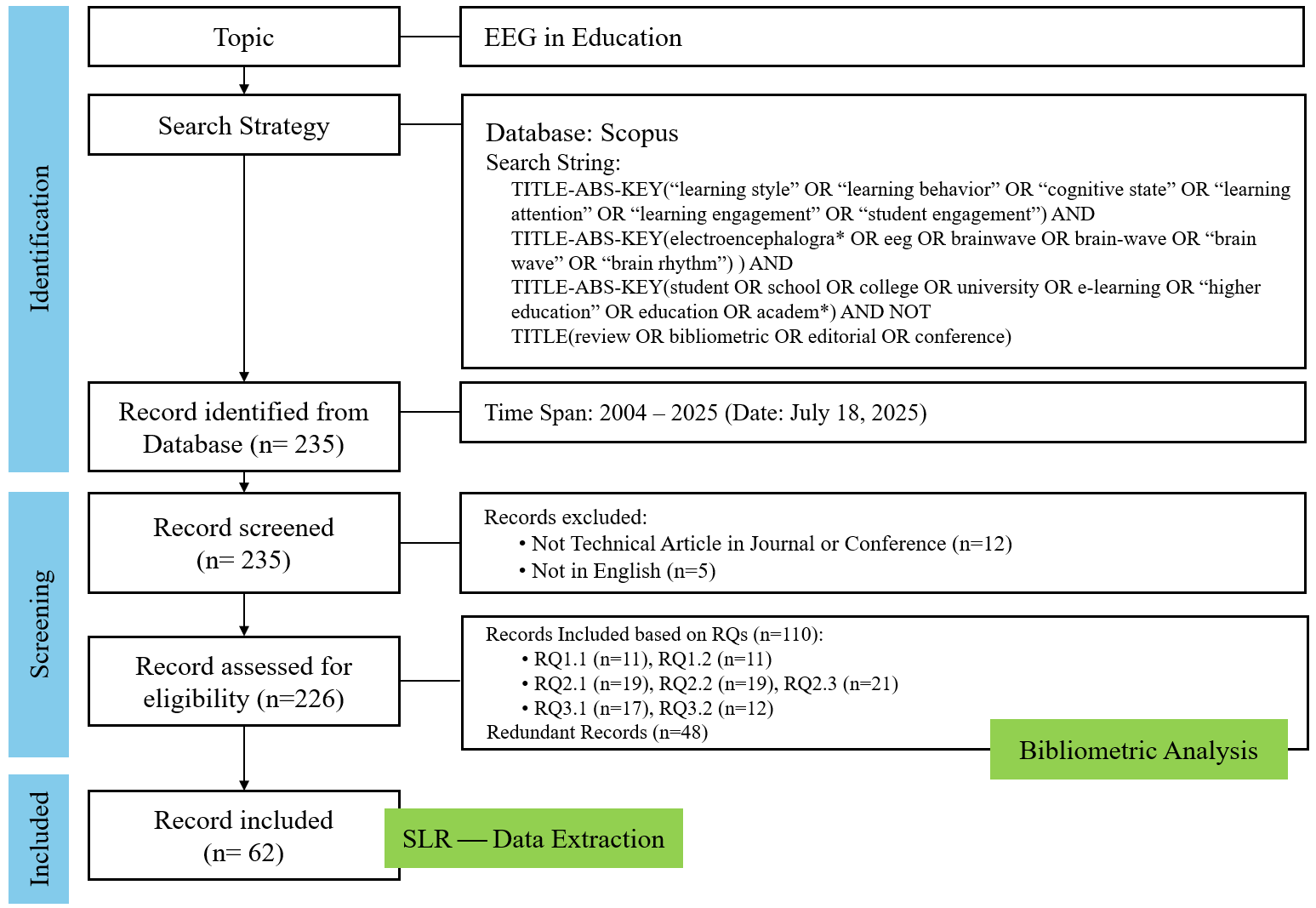}
    \caption{Adapted PRISMA flow diagram}
    \label{fig:prisma}
\end{figure}

\subsection{BenSLR: The Fusion between BA and SLR}
As illustrated in Figure 2, the BenSLR (Bibliometric-enhanced Systematic Literature Review) approach integrates the standard SLR process with Bibliometric Analysis (BA) to refine and augment the development of RQs. Even though it follows existing Systematic Literature Review (SLR) procedures \parencite{gough2017introduction,kitchenham2015evidence}, BenSLR brings a bibliometric element—the application of co-occurrence analysis—to the search strategy to uncover important areas of interest. These areas serve a quantitative outline for the creation of more complex RQs, serving directly to inform study selection, quality assessment, data extraction, and synthesis. Compared to existing SLR approaches, which often rely largely on manual screening and qualitative synthesis, BenSLR draws upon BA to uncover hidden patterns, associations, and trends within the literature \parencite{marziGuidelinesBibliometricSystematicLiterature2025,silvaStartupsValuationBibliometric2021}, thus enabling RQs to be crafted with improved accuracy and an evidence-based grounding. The combination of SLR and BA thus makes BenSLR an innovative methodological framework, granting both detailed and macroscopic insights into the evolution and organization of a research field.\\

\begin{figure}[h]
    \centering
    \includegraphics[width=0.6\textwidth]{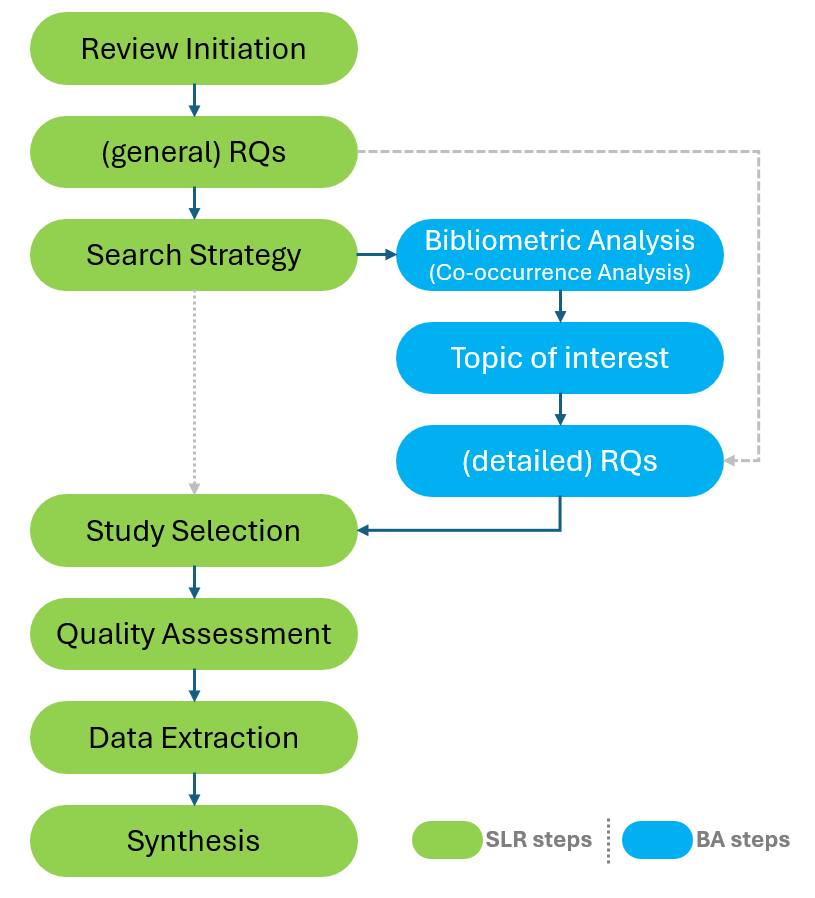}
    \caption{BenSLR steps}
    \label{fig:benslr}
\end{figure}

The combination of BA within the SLR creates an embedded and iterative process, rather than a comparison of quantitative and qualitative methods. Bibliometric analysis, covering performance analysis, science mapping, and network analysis, enables the possibility of identifying key publications, influential authors, institutions, and thematic groups \parencite{mukhtarSystematicLiteratureReview2025,sharmaSystematicLiteratureReview2024}. Outputs from bibliometric analysis are then employed to guide the systematic review by identifying influential studies and emerging patterns to yield a more unified and data-informed narrowing of research questions (RQs) \parencite{marziGuidelinesBibliometricSystematicLiterature2025}. Positioning BA as a central component, BenSLR extends beyond the usual confines of SLRs, providing a distinct and methodological avenue from broad inquiry to narrow, high-quality output. This combined approach not only fortifies method strength but also allows a dynamic framework of advancing theories and an agenda for follow-up research, thereby illustrating the unique contribution and novelty of BenSLR to refining systematic literature review approaches.\\

\subsection{Study Selection and Quality Assessment}
The selection of studies was based on a systematic screening protocol in accordance with the PRISMA guidelines. Step 1 included the identification of probable relevant studies from metadata scanning. Abstracts and titles were screened for relevance to the respective RQs (RQ1.1, RQ1.2, RQ2.1, RQ2.2, RQ2.3, RQ3.1, and RQ3.2), and the records were included if there was an evident connection to at least a single RQ. Studies not including EEG-based education research or methodologically irrelevant ones were excluded in this step.\\

After the step of selection, full-text copies of the selected studies were obtained for quality assessment. This step was useful in ensuring that not only did the studies align with the research questions on the metadata level, but also effectively discussed the pertinent dimensions of research. Each article was duly examined to determine if its methodologies, results, or discussions served to help adequately answer its respective research question. Articles lacking in detail, methodological strength, or clear association with the themes of the research questions did not pass. Only those articles carrying sufficient and relevant evidence made it to the final synthesis.\\

\subsection{Data Extraction and Synthesis}
Under the BenSLR protocol, the data extraction happens after refining initial research questions to well-articulated and clear-specified aims. For any study satisfying eligibility requirements, information is carefully extracted following the content relevant to the refined research questions, including methodological approaches, computational techniques, EEG signal processing procedures, educational constructs, cognitive and emotional experiences, and applications within innovative instructional technologies. A complete overview of the results of all studies is presented in a systematic table structure containing aggregate data. This tabular organization ensures the process of data collection is transparent, reliable, and directly aligns with the aims of the review, though the specific sub-questions are to be stated in the latter part of the Results section. After data extraction, synthesis is performed using an interpretation of the findings organized to answer the research questions. Both quantitative patterns, such as frequency of use or method of compilation, and qualitative information obtained from the content of the study are combined to identify the main methods used, emerging trends, and areas that have not been covered. By aligning the organized evidence to the distilled research questions, the BenSLR methodology ensures the synthesis is systematic, evidence-informed, and analytically rigorous, and creates a harmonious framework for the inclusive results outlined in the following section.\\

\section{Results}
\subsection{Co-occurrence analysis results}
To establish the thematic outline of EEG-centric learning studies, a VOSviewer version 1.6.20 \parencite{vanEck2010} co-occurrence analysis of author keywords was performed. A thesaurus file was utilized to ensure consistency in the mapping procedure so that keywords with identical connotations are represented by a single standard term. Redundancy was eliminated through this process, and the precision of the visualization improved, ensuring the resultant clusters depict genuine conceptual relationships rather than differences in terminology.\\

Figure 3 shows the network visualization of the co-occurrence analysis, which identified three distinct clusters of keywords encapsulating the thematic structure of EEG-based educational research. Each of the clusters is a group of interconnected concepts that tend to coalesce within the body of work and thus highlights methodological, cognitive, and applied dimensions of the field. The clustering showcases the development of the body of work in this direction into both interrelated and disparate themes and thus provides a data-informed picture of the landscape of knowledge. This bibliometric viewpoint provides a systematic foundation for refining the focus of inquiry, as it spans both entrenched areas and emerging pathways within the body of work.\\

Under the BenSLR framework, the extracted clusters not only described the field but also played a fundamental part in the sharpening of the research questions. The subjects of interest extracted from each of the clusters were matched against the above study's overarching inquiry to give rise to the establishment of detailed sub-questions for RQ1, RQ2, and RQ3. Specifically, methodological and computational themes helped in the articulation of RQ1, cognitive and affective constructions in RQ2, and subjects concerned with trends or voids helped provide an advancement of RQ3. Through the incorporation of the co-occurrence analysis into the process of reviewing, BenSLR ensures that the research questions are backed with bibliometric evidence, thereby enhancing their precision, pertinence, and alignment with the true structure of the field.\\

\begin{figure}[h]
    \centering
    \includegraphics[width=0.8\textwidth]{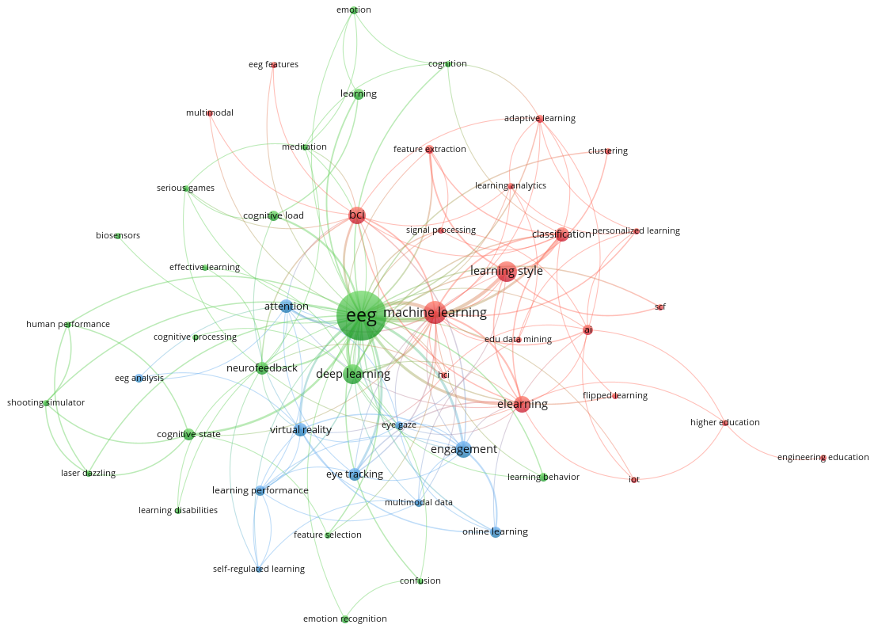}
    \caption{Network visualization from co-occurrence analysis resulted in 3 clusters}
    \label{fig:occurrence}
\end{figure}

\subsection{Paper Collection}
The screening process yielded a varying number of eligible papers across the refined research questions. There are 22 studies related to RQ1, 11 studies related to RQ1.1 discussing computational techniques, while the other 11 are related to RQ1.2 discussing EEG signal processing approaches. In RQ2, 19 studies examined cognitive and affective states (RQ2.1), 19 studies explored individual differences and learning outcomes (RQ2.2), and 21 studies focused on the application of EEG in the context of education and neurofeedback (RQ2.3). For RQ3, 17 studies discussed thematic and methodological trends (RQ3.1), while 12 studies highlighted research gaps and future directions (RQ3.2). Even after many publications covering several research questions, the subsequent corpus included 62 different studies. Redundancy here indicates that many studies within the field of EEG-informing educational research include different dimensions and simultaneously touch on methodological, cognitive, and applied sides. These 62 different articles provide a foundational basis for the subsequent data synthesis and extraction and provide a balanced yet comprehensive evidentiary groundwork for tackling the research questions.\\

\subsection{RQ1 Detailed Results}
A co-occurrence analysis was conducted on the corpus of EEG-related education studies, as can be seen from Figure 4, with an aim to detect keywords and thematic clusters that tend to co-occur. From the analysis, two main sets of aspects of interest emerged: the first set—machine learning, deep learning, and classification—represents the computational methodologies discussed within the body of work, while the second set—signal processing, EEG analysis, feature extraction, and feature selection—indicates the basic steps involved in the preprocessing of EEG signals for subsequent analysis. The extracted aspects of interest then guided the development of more detailed research questions, particularly RQ1.1 and RQ1.2, for the purposes of conducting an in-depth investigation of methodological approaches utilized within EEG-based education research.\\

\begin{figure}[h]
    \centering
    \includegraphics[width=0.8\textwidth]{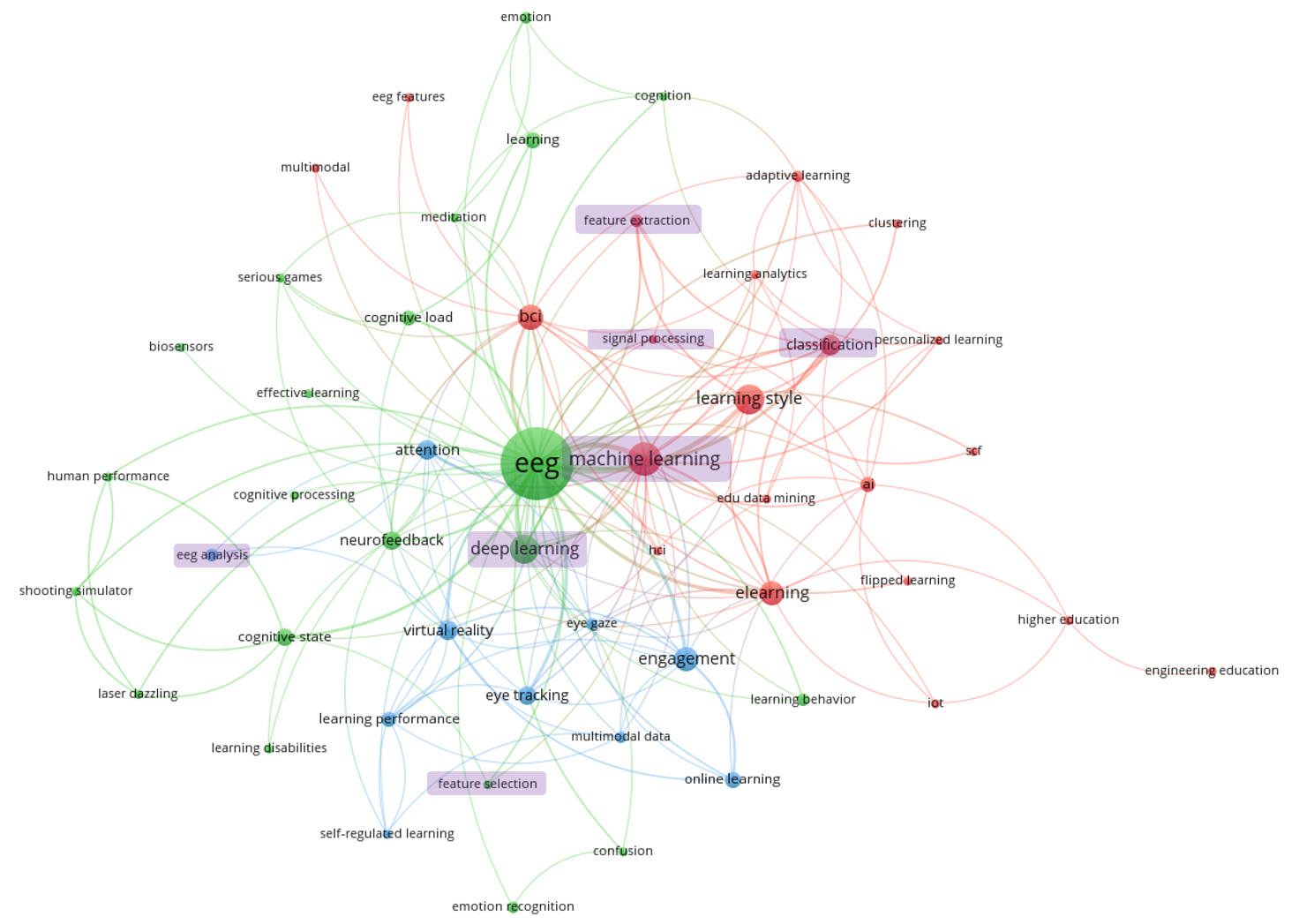}
    \caption{Topic of interest for RQ1}
    \label{fig:topic}
\end{figure}

The first group of subjects of interest—the machine learning, deep learning, and classification—is relevant to RQ1.1: What computational approaches are employed in EEG-based learning studies, and what has been the history of their development? As Table 1 indicates, machine learning approaches are widely applied across studies to predict student engagement, cognitive states, and learning patterns from EEG measures. Deep learning techniques are increasingly employed to capture complex spatiotemporal patterns embedded in EEG measures and provide automatic feature extraction. Classification approaches are employed to map EEG measures to cognitive and learning results, including confusion, attentiveness, learning styles, working memory, and emotion/engagement patterns. For algorithms and complete study details, see Table 1.\\

The second group of subjects, comprising signal processing, EEG analysis, feature extraction, and feature selection, pertains to RQ1.2: How is the EEG signal processed and made amenable to meaningful educational evaluation? EEG signals are made ready to eliminate noise and artifacts and ensure data purity for further analysis. Feature extraction converts signals to meaningful ones using statistical, spectral, and transform-based measures. Feature selection methods are employed for determining the most relevant features to be utilized for cognitive modelling of learners. For study-level and method-level details, refer to Table 2.\\
\begin{table}[H]
  \centering
  \caption{Computational approaches and classification in EEG-based educational research}
  \label{tab:comp}
  \renewcommand{\arraystretch}{1.05}
  \setlist[itemize]{nosep, leftmargin=*}
  
  \begin{tabularx}{\textwidth}{|>{\raggedright\arraybackslash}p{3.2cm}|>{\raggedright\arraybackslash}p{3.2cm}|Y|}
  \hline
  \textbf{Topic} & \textbf{Aspect} & \textbf{Findings} \\
  \hline
  \textbf{Machine Learning} & Algorithms &
  \begin{itemize}
    \small
    \item Decision Tree \parencite{miahODLBCIOptimalDeep2024}
    \item AdaBoost \parencite{miahODLBCIOptimalDeep2024}
    \item Bagging \parencite{chowdhuriRealtimeClassificationEEG2024,miahODLBCIOptimalDeep2024}
    \item MLP \parencite{miahODLBCIOptimalDeep2024,yuvarajMachineLearningFramework2024a}
    \item Naive Bayes \parencite{chowdhuriRealtimeClassificationEEG2024,miahODLBCIOptimalDeep2024}
    \item Random Forest \parencite{chenApplicationElectroencephalographySensors2024,chowdhuriRealtimeClassificationEEG2024,dursoEnhancingEducationalOutcomes2024b,elkerdawyBuildingCognitiveProfiles2020,miahODLBCIOptimalDeep2024,yuvarajMachineLearningFramework2024a}
    \item SVM \parencite{chowdhuriRealtimeClassificationEEG2024,dursoEnhancingEducationalOutcomes2024b,elkerdawyBuildingCognitiveProfiles2020,miahODLBCIOptimalDeep2024}
    \item XGBoost \parencite{miahODLBCIOptimalDeep2024}
    \item Logistic Regression \parencite{elkerdawyBuildingCognitiveProfiles2020}
    \item Neural Networks \parencite{dursoEnhancingEducationalOutcomes2024b,elkerdawyBuildingCognitiveProfiles2020}
    \item KNN \parencite{dursoEnhancingEducationalOutcomes2024b}
    \item Multivariate Linear Regression \parencite{chenApplicationElectroencephalographySensors2024}
  \end{itemize}
  \\
  \hline
  
  \textbf{Deep Learning} & Architectures &
  \begin{itemize}
    \small
    \item CNN, RNN \parencite{chenApplicationElectroencephalographySensors2024,miahODLBCIOptimalDeep2024}
    \item LSTM, FCNN, ConvLSTM \parencite{elkerdawyBuildingCognitiveProfiles2020,yuvarajMachineLearningFramework2024a}
    \item LSTM-CNN, LSTM-FCNN \parencite{jawedDeepLearningBasedAssessment2024b}
    \item Transformer-based models \parencite{elgammalNeuroTutorNeuralDecoding2025}
  \end{itemize}
  \\
  \hline
  
  \textbf{Classification} & Cognitive \& learning outcomes &
  \begin{itemize}
    \small
    \item Confusion \parencite{miahODLBCIOptimalDeep2024}
    \item Attentiveness / Engagement \parencite{chowdhuriRealtimeClassificationEEG2024,dursoEnhancingEducationalOutcomes2024b}
    \item Learning styles \parencite{jawedDeepLearningBasedAssessment2024b,yuvarajMachineLearningFramework2024a}
    \item Basic cognitive states / working memory / attention \parencite{elkerdawyBuildingCognitiveProfiles2020}
    \item Attention focus / memory / cognitive performance \parencite{chenApplicationElectroencephalographySensors2024}
    \item Emotion \& engagement patterns \parencite{elgammalNeuroTutorNeuralDecoding2025}
  \end{itemize}
  \\
  \hline
  \end{tabularx}
\end{table}

\begin{table}[H]
  \centering
  \caption{EEG signal processing, feature extraction, and selection in educational research}
  \label{tab:eeg}
  \renewcommand{\arraystretch}{1.05}
  
  \begin{tabularx}{\textwidth}{|>{\raggedright\arraybackslash}p{4cm}|>{\raggedright\arraybackslash}p{3.5cm}|Y|}
  \hline
  \textbf{Topic} & \textbf{Aspect} & \textbf{Findings} \\
  \hline
  
  \textbf{Signal Acquisition \& Preprocessing} & Devices \& Filtering &
  \begin{itemize}\small
    \item EEG collected via Emotiv headsets \parencite{arnaldoComputerizedBrainInterfaces2018,mohamedFacilitatingClassroomOrchestration2020} and 64-channel systems \parencite{liuLearnerCognitiveState2024}
    \item Filtering: bandpass, notch, low-pass filters \parencite{chowdhuriRealtimeClassificationEEG2024,elgammalNeuroTutorNeuralDecoding2025,jawedDeepLearningBasedAssessment2024b,yuvarajMachineLearningFramework2024a}
    \item Artifact removal via ICA or automated rejection \parencite{jawedDeepLearningBasedAssessment2024b,liuLearnerCognitiveState2024,yuvarajMachineLearningFramework2024a}
  \end{itemize}
  \\
  \hline
  
  \textbf{EEG Analysis} & Frequency \& Coherence &
  \begin{itemize}\small
    \item Time-frequency analysis via FFT / PSD \parencite{chowdhuriRealtimeClassificationEEG2024,liBiosensorTechnologyAdaptive2025a,sulaimanDevelopmentEEGBasedSystem2022}
    \item Standard bands alpha, beta, theta, delta, gamma \parencite{chowdhuriRealtimeClassificationEEG2024,dursoEnhancingEducationalOutcomes2024b,sulaimanDevelopmentEEGBasedSystem2022}
    \item Coherence / inter-regional coordination \parencite{dursoEnhancingEducationalOutcomes2024b}
  \end{itemize}
  \\
  \hline
  
  \textbf{Feature Extraction} & Metrics \& Indices &
  \begin{itemize}\small
    \item Statistical features (mean, SD, skewness, kurtosis) \parencite{liuLearnerCognitiveState2024}
    \item Hjorth parameters, fractal dimension, higher-order spectra \parencite{elgammalNeuroTutorNeuralDecoding2025,jawedDeepLearningBasedAssessment2024b}
    \item Power Spectral Entropy, DWT coefficients \parencite{yuvarajMachineLearningFramework2024a}
    \item Cognitive indices \parencite{dursoEnhancingEducationalOutcomes2024b}
  \end{itemize}
  \\
  \hline
  
  \textbf{Feature Selection} & Dimensionality \& Optimization &
  \begin{itemize}\small
    \item PCA \parencite{liuLearnerCognitiveState2024,yuvarajMachineLearningFramework2024a}
    \item ANOVA \parencite{elgammalNeuroTutorNeuralDecoding2025,jawedDeepLearningBasedAssessment2024b}
    \item Sequence Forward Selection \parencite{jawedDeepLearningBasedAssessment2024b}
    \item Genetic Algorithm + Random Forest \parencite{elgammalNeuroTutorNeuralDecoding2025,yuvarajMachineLearningFramework2024a}
    \item Feature evaluation to identify discriminatory features \parencite{elgammalNeuroTutorNeuralDecoding2025}
  \end{itemize}
  \\
  \hline
  
  \end{tabularx}
\end{table}

The co-occurrence plot in Figure 4 illustrates how the themes of existing literature led to the formulation of Research Questions RQ1.1 and RQ1.2. The machine learning, deep learning, and classification aspects (RQ1.1) refer to the computational methodologies utilized, while signal processing, feature extraction, and feature selection aspects (RQ1.2) refer to the stepwise procedures entailed in EEG signal preparation. These combined tailored insights provide a comprehensive overview of the prevalent methodological approaches within EEG-focused educationally directed research.\\

\subsection{RQ2 Detailed Results}
Following the approach used for the detailed analysis of RQ1, RQ2 was similarly examined by considering several topics of interest identified from the network visualization of the co-occurrence analysis, as shown in Figure 5. From this analysis, three sets of topics of interest emerged: the first set—attention, cognitive state, engagement, and learning behavior—reflects the cognitive and affective constructs frequently explored in the literature; the second set—learning performance, personalized learning, and learning styles—represents investigations into individual differences and learning outcomes; and the third set—virtual reality and neurofeedback—highlights the emerging applications of EEG in adaptive and immersive educational environments. These topics of interest informed the formulation of more specific research questions, namely RQ2.1, RQ2.2, and RQ2.3, to guide a detailed examination of educational constructs and learning-related phenomena in EEG-based studies.\\

\begin{figure}[h]
    \centering
    \includegraphics[width=1\textwidth]{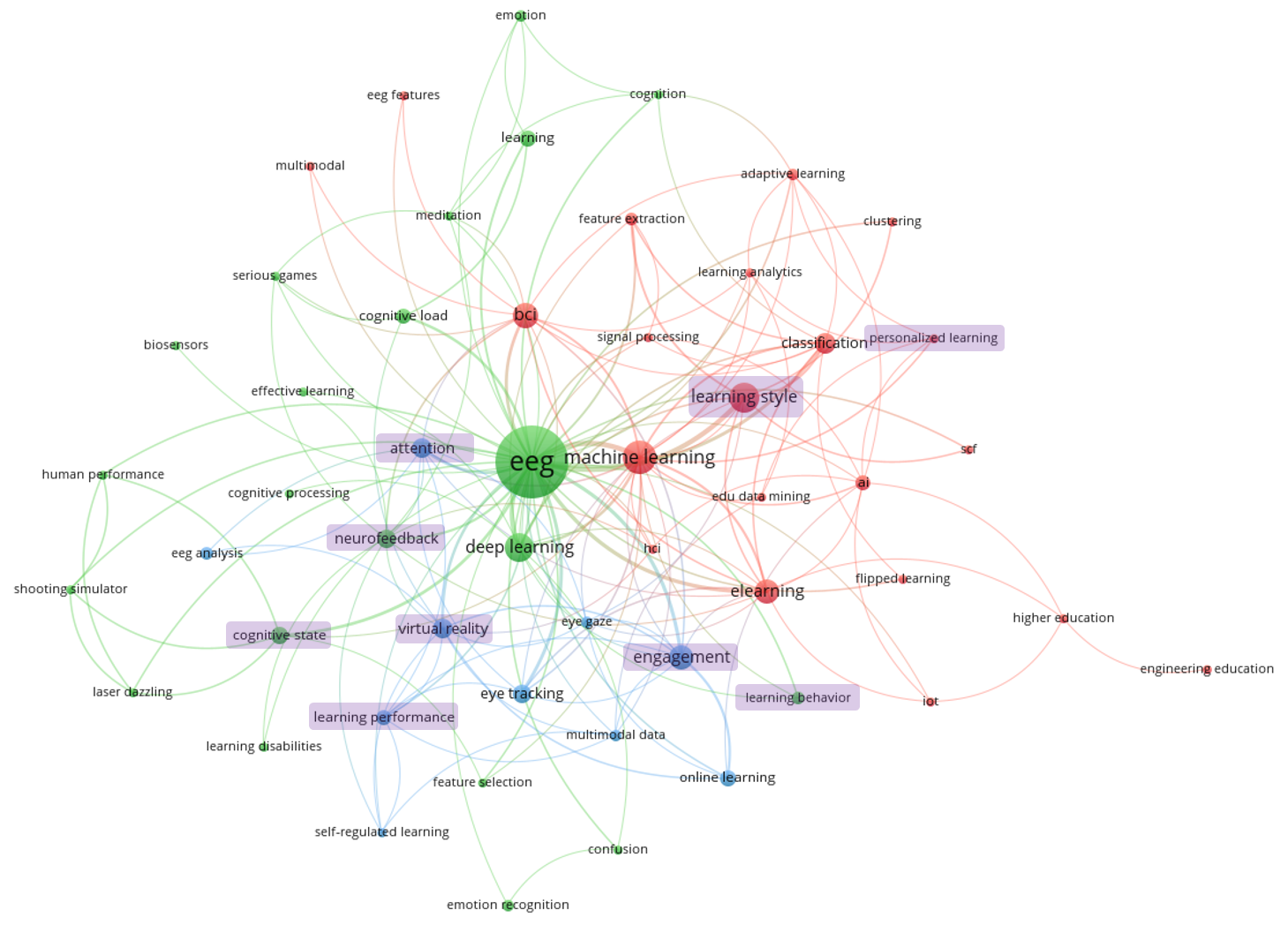}
    \caption{Topic of Interest for RQ2}
    \label{fig:topicrq2}
\end{figure}

The next four topics of interest, namely, attention, cognitive state, engagement, and learning behavior, are related to RQ2.1: What cognitive and affective states are commonly explored in EEG-based educational studies? The use of EEG, as summarized in Table 3, has been quite diverse in the case of attention. Different attention types were covered by the studies, including focused, sustained, selective, divided, and alternating attention, as well as vigilance and attentional lapses. The cognitive states that were examined most often were those that accompanied working memory, cognitive workload, problem-solving, reasoning, planning, and overall mental effort. These states were evaluated so that learners’ task demands and learning processes could be identified. Engagement can be seen as the integration of cognitive and emotional aspects, and it is gauged through such constructs as focus, involvement, flow, and motivation, while affective states, e.g., stress, fatigue, and arousal, are leading factors in the investigation of their impact on learning outcomes. Besides memory performance, learning behavior was also looked at for the purpose of monitoring the changes (i.e., learning styles, learning preferences, and adaptive responses) in the learners’ interaction with instructional materials. For the details of the studies and the exact EEG measures, see Table 3.\\

\begin{longtable}{|p{3.5cm}|p{4cm}|p{7.5cm}|}
  \caption{Cognitive and affective states in EEG-based educational research}
  \label{tab:cognitive} \\
  \hline
  \textbf{Topic} & \textbf{Aspect} & \textbf{Findings} \\
  \hline
  \endfirsthead

  \multicolumn{3}{c}%
  {\tablename\ \thetable\ -- \textit{continued from previous page}} \\
  \hline
  \textbf{Topic} & \textbf{Aspect} & \textbf{Findings} \\
  \hline
  \endhead

  \hline \multicolumn{3}{|r|}{\textit{Continued on next page}} \\ \hline
  \endfoot

  \hline
  \endlastfoot

\textbf{Attention} & Focused \& Sustained Attention &
\begin{itemize}[noitemsep, topsep=0pt, parsep=0pt, partopsep=0pt, left=1.2em]\small
  \item Focused attention, sustained attention, selective attention, divided attention, alternating attention \parencite{chiangEEGBasedDetectionModel2018}
  \item Vigilance sebagai sustained attention \parencite{binabdulrashidDeterminationVigilanceboundLearning2012}
  \item Visual attention \parencite{grammerEffectsContextNeural2021a}
  \item Attentive vs. inattentive, four attention states \parencite{mohamedFacilitatingClassroomOrchestration2020}
  \item Attention lapses, shared attention \parencite{waznyRealTimeCognitiveStateNeuroimaging2018}
  \item Attention differences by media type \parencite{niEEGBasedAttentionAnalysis2020a}
  \item Instantaneous attention \parencite{elkerdawyBuildingCognitiveProfiles2020}
  \item Concentrated attention \parencite{apicellaEEGbasedMeasurementSystem2022}
  \item Attention in online learning \parencite{talukdarOnEEEGbasedPassive2023}
\end{itemize}
\\ \hline

\textbf{Cognitive State} & Working Memory \& Cognitive Load &
\begin{itemize}[noitemsep, topsep=0pt, parsep=0pt, partopsep=0pt, left=1.2em]\small
  \item Working memory load \parencite{cheahaEEGNeuralSubstrates2024,mohamedFacilitatingClassroomOrchestration2020}
  \item Cognitive workload \parencite{cheahaEEGNeuralSubstrates2024,duAnalyzingEffectsInstructional2024b,goldbergEfficacyMeasuringEngagement2012,waznyRealTimeCognitiveStateNeuroimaging2018}
  \item Recognition/memory performance \parencite{mohamedFacilitatingClassroomOrchestration2020}
  \item Episodic memory, memory retrieval, spatial cognition \parencite{cheahaEEGNeuralSubstrates2024}
  \item Perception and visual encoding \parencite{elkerdawyBuildingCognitiveProfiles2020,grammerEffectsContextNeural2021a}
  \item Fatigue \parencite{cheahaEEGNeuralSubstrates2024,dursoEnhancingEducationalOutcomes2024b}
  \item Stress \parencite{dursoEnhancingEducationalOutcomes2024b}
\end{itemize}
\\ \hline

\textbf{Engagement} & Emotional \& Cognitive Engagement &
\begin{itemize}[noitemsep, topsep=0pt, parsep=0pt, partopsep=0pt, left=1.2em]\small
  \item Cognitive engagement \parencite{apicellaEEGbasedMeasurementSystem2022,cheahaEEGNeuralSubstrates2024,davidescoDetectingFluctuationsStudent2023}
  \item Emotional engagement \parencite{apicellaEEGbasedMeasurementSystem2022,sandhuEvaluationLearningPerformance2017}
  \item Behavioral engagement \parencite{apicellaEEGbasedMeasurementSystem2022}
  \item General engagement indices from EEG oscillations \parencite{christoforouYourBrainSTEM2024,goldbergEfficacyMeasuringEngagement2012,grammerEffectsContextNeural2021a,natalizioRealtimeEstimationEEGbased2024}
  \item Engagement via frontal asymmetry \parencite{sandhuEvaluationLearningPerformance2017}
  \item Engagement tracking in virtual environments \parencite{natalizioRealtimeEstimationEEGbased2024}
\end{itemize}
\\ \hline

\textbf{Learning Behavior} & Learning Performance \& Adaptive Responses &
\begin{itemize}[noitemsep, topsep=0pt, parsep=0pt, partopsep=0pt, left=1.2em]\small
  \item Learning style detection \parencite{binabdulrashidDeterminationVigilanceboundLearning2012,deenadayalanEEGBasedLearners2018a,niEEGBasedAttentionAnalysis2020a}
  \item Learning behavior prediction \parencite{niEEGBasedAttentionAnalysis2020a,waznyRealTimeCognitiveStateNeuroimaging2018}
  \item Confusion/brain fog as a learning barrier \parencite{talukdarOnEEEGbasedPassive2023}
  \item Retention and memory \parencite{davidescoDetectingFluctuationsStudent2023}
  \item Cognitive load effect on learning \parencite{apicellaEEGbasedMeasurementSystem2022,duAnalyzingEffectsInstructional2024b}
  \item Adaptive/personalized responses \parencite{elkerdawyBuildingCognitiveProfiles2020,goldbergEfficacyMeasuringEngagement2012,mohamedFacilitatingClassroomOrchestration2020}
\end{itemize}
\\ \hline

\end{longtable}

The topics in the second group—namely learning performance, personalized learning, and learning styles—are related to RQ2.2: How is EEG utilized to assess individual differences and learning outcomes? As indicated in Table 4, the main goal of using EEG is to determine the attention of the learner, the cognitive load, and the efficiency of the task, where a few metrics, such as alpha, beta, and theta power spectral density (PSD), may reflect engagement, flow state, and problem-solving performance. Besides this, brain activity monitoring with EEG is also helpful in differentiating between the different dimensions of the Felder–Silverman model and active–reflective preferences during the assessment of the learning style, while some other methods are at work for the identification of non-FSLSM distinctions like visual versus non-visual learners. These objective EEG-based measures can uncover individual cognitive profiles and learning dynamics that can result in the development of adaptive and personalized educational interventions that not only consider the factors of attention, engagement, and learning style but also do not treat them as separate constructs.\\

\begin{longtable}{|p{3.5cm}|p{4.2cm}|p{7.8cm}|}
  \caption{EEG-based educational research topics, aspects, and findings}
  \label{tab:EEG-topics} \\
  \hline
  \textbf{Topic} & \textbf{Aspect} & \textbf{Findings} \\
  \hline
  \endfirsthead
  
  \multicolumn{3}{c}%
  {\tablename\ \thetable\ -- \textit{continued from previous page}} \\
  \hline
  \textbf{Topic} & \textbf{Aspect} & \textbf{Findings} \\
  \hline
  \endhead
  
  \hline \multicolumn{3}{|r|}{\textit{Continued on next page}} \\ \hline
  \endfoot
  
  \hline
  \endlastfoot
  
  \textbf{Learning performance} & 
  Attention, cognitive strategies, and task performance &
  \begin{itemize}\small
    \item Attention levels (alpha, beta, theta PSD) indicate learning efficiency \parencite{chiangEEGBasedDetectionModel2018,niEEGBasedAttentionAnalysis2020a}
    \item Flow state and engagement linked to academic performance \parencite{dursoEnhancingEducationalOutcomes2024b}
    \item Brain activity correlates with problem-solving performance \parencite{cordovaIdentifyingProblemSolving2015}
    \item EEG-measured attention predicts learning outcomes in multimedia learning \parencite{niEEGBasedAttentionAnalysis2020a}
  \end{itemize}
  \\ \hline
  
  \textbf{Personalized learning} &
  Tailoring learning content based on EEG-derived individual differences &
  \begin{itemize}\small
    \item Adaptive learning content adjusted by cognitive state and working memory load \parencite{arnaldoComputerizedBrainInterfaces2018}
    \item Real-time EEG feedback informs student engagement strategies \parencite{dursoEnhancingEducationalOutcomes2024b}
    \item EEG-informed adjustments in e-learning environments based on emotional state and attention \parencite{alhasanExperimentalStudyLearning2018}
    \item Personalized teaching strategies informed by gender- and style-specific brain activation patterns \parencite{hamesEEGbasedComparisonsPerformance2013}
  \end{itemize}
  \\ \hline
  
  \textbf{Learning styles} &
  Classification of learners based on EEG patterns &
  \begin{itemize}\small
    \item EEG-based detection of Felder–Silverman learning styles, including active–reflective dimension \parencite{hasibuanProposedModelDetecting2025b,zhangDesignImplementationEEGBased2021h,zhangTSMGDeepLearning2021e}
    \item EEG-based identification of active vs. reflective learners (non–Felder–Silverman) \parencite{niEEGBasedAttentionAnalysis2020a}
    \item Kolb’s four learning styles identified via alpha and theta sub-bands, spectral features \parencite{binabdulrashidAppraisalEEGBeta2015a,rashid2014summative,megataliLearningStyleClassification2014d,megataliEEGSUBBANDSPECTRAL2016d}
    \item Visual vs. non-visual learners distinguished via PSD and DWT features \parencite{jawedDeepLearningBasedAssessment2024b}
    \item EEG signals correlated with personality traits influencing learning style \parencite{rashid2014summative}
  \end{itemize}
  \\ \hline
  
  \end{longtable}

The third group of topics – virtual reality and neurofeedback – is related to RQ2.3: How is EEG integrated into emerging educational environments and interventions? As can be seen from the summary in Table 5, the EEG is used to measure the cognitive and emotional states of the users by monitoring attention, memory, emotions, and cognitive load via a scientific technique called brain recording. In virtual reality settings, the EEG data are used to adapt the avatars, the difficulty of the task, and the different parts of the simulation, and thus, the students can have their own immersive and personalized learning environments. Neurofeedback programs are designed to use the brain signals (EEG) to give immediate feedback that enables the learners to self-control, concentrate more, and acquire metacognitive skills. Portable and low-cost EEG devices, which allow these interventions to be implemented in both classroom and online settings, have been released into the market. These devices support the dynamic and interactive feedback loops that facilitate engagement and optimize educational outcomes. Details of the study level and aspects can be found in Table 5.\\

\renewcommand{\arraystretch}{1.2}

\begin{longtable}{|>{\raggedright\arraybackslash}p{3.5cm}|
                        >{\raggedright\arraybackslash}p{4cm}|
                        >{\raggedright\arraybackslash}p{8cm}|}
  \caption{EEG integration in emerging educational environments and interventions}
  \label{tab:EEG-VR-Neurofeedback} \\
  \hline
  \textbf{Topic} & \textbf{Aspect} & \textbf{Findings} \\
  \hline
  \endfirsthead
  
  \multicolumn{3}{c}%
  {\tablename\ \thetable\ -- \textit{continued from previous page}} \\
  \hline
  \textbf{Topic} & \textbf{Aspect} & \textbf{Findings} \\
  \hline
  \endhead
  
  \hline \multicolumn{3}{|r|}{\textit{Continued on next page}} \\ \hline
  \endfoot
  
  \hline
  \endlastfoot
  
  \multirow[t]{6}{*}{\raggedright\textbf{Virtual Reality}} 
& \multirow[t]{6}{*}{\raggedright Assessment} &
  \begin{itemize}\small
    \item Real-time EEG-based assessment of attention, memory, and emotional states in adaptive VR environments \parencite{hubbardEnhancingLearningVirtual2017}
    \item EEG monitoring of cognitive workload during VR tasks \parencite{tremmelEstimatingCognitiveWorkload2019a}
    \item EEG measurement of memory encoding and cognitive load influenced by pedagogical agents and visual cues \parencite{jiCanPedagogicalAgents2025}
    \item EEG-based engagement and attention assessment in VR classrooms, combined with eye-tracking \parencite{sivaprasadDoctoralColloquiumUtilizing2022}
    \item Passive EEG monitoring of workload, intuition, and relaxation embedded in VR headsets \parencite{leeEEGbasedEvaluationIntuitive2022}
    \item EEG assessment of attention in remote VR learning environments \parencite{wangEvaluateLearningAttention2022}
  \end{itemize}
  \\ \hline
  
  \multirow[t]{6}{*}{\raggedright\textbf{Neurofeedback}} 
  & \multirow[t]{6}{*}{\raggedright Intervention} &
  \begin{itemize}\small
    \item EEG-based neurofeedback to enhance algorithmic thinking and cognitive activation \parencite{plerouEEGAnalysisNeurofeedback2017}
    \item Game-based EEG neurofeedback for sustained attention and executive function \parencite{krellSchoolBasedNeurofeedbackTraining2023} 
    \item EEG-driven attention deficit detection with real-time visual/audio feedback for personalized learning \parencite{handayaniConceptPaperAlert2024}
    \item Neurofeedback via BCI and eye-tracking for online learning engagement \parencite{jamilImprovingStudentsCognitive2023}
    \item EEG-based biofeedback for meditation, relaxation, and cognitive skill training in serious games \parencite{dursoEnhancingEducationalOutcomes2024b,limLearningAttentionImprovement2018,murdochExperientialLearningBasedApproach2019}
    \item Classification of neural responses in adaptive virtual and robot tutors for neurofeedback-informed learning \parencite{elgammalNeuroTutorNeuralDecoding2025}
  \end{itemize}
  \\ \hline
  
\end{longtable}

The co-occurrence analysis overall in Figure 5 depicts the connection between the themes that were identified and the development of RQ2.1, RQ2.2, and RQ2.3. RQ2.1, which deals with attention, cognitive conditions, engagement, and learning behavior, is among those topics that encompass very important cognitive and affective constructs. RQ2.2 discusses learning performance, personalized learning, and learning styles with a focus on individual differences and learning evaluation. RQ2.3 describes virtual reality and neurofeedback, with some new applications of EEG for adaptive and immersive educational environments. The paper shows the delineation of the construct, learner characteristics, and innovative interventions as the main themes of the educational research based on the use of EEG.\\

\subsection{RQ3 Detailed Results}
For the sake of thoroughly investigating RQ3—experimental trends and future possibilities about Electroencephalogram (EEG) in education, the topics of interest, based on the Enhanced Strategic Diagram, were analyzed in a more detailed and structured manner. The topics' characteristics of each quadrant of the thematic map—were examined from thematic and methodological perspectives—RQ3.1—what thematic and methodological trends have shaped the development of EEG-based research in education? —was the first question that resulted from the first group of topics: machine learning, learning style, deep learning, EEG, engagement, and attention. Motivational constructs such as engagement and attention, presently also commonly studied due to their impact on learning outcomes, still lie at the center of conceptual frameworks used \parencite{chowdhuriRealtimeClassificationEEG2024,elkerdawyBuildingCognitiveProfiles2020,sandhuEvaluationLearningPerformance2017}. Besides electrophysiological correlates of attention and engagement, learning styles have also been explored to better understand how instructional strategies impact learning neural correlates in both laboratory and classroom settings \parencite{grammerEffectsContextNeural2021a,niEEGBasedAttentionAnalysis2020a}. Some examples of methodological trends include the use of portable EEG devices for naturalistic data collection and computational approaches such as machine learning and deep learning, including SVM, Random Forests, CNNs, and LSTMs for modeling complex EEG patterns and classifying cognitive states \parencite{apicellaEEGbasedMeasurementSystem2022,chenApplicationElectroencephalographySensors2024,dursoEnhancingEducationalOutcomes2024b}. Multimodal integration with eye-tracking or facial expression data significantly improves both the ecological validity and granularity of findings \parencite{christoforouYourBrainSTEM2024,zhangMultimodalFastSlow2023}.\\

\begin{figure}[h]
  \centering
  \includegraphics[width=1\textwidth]{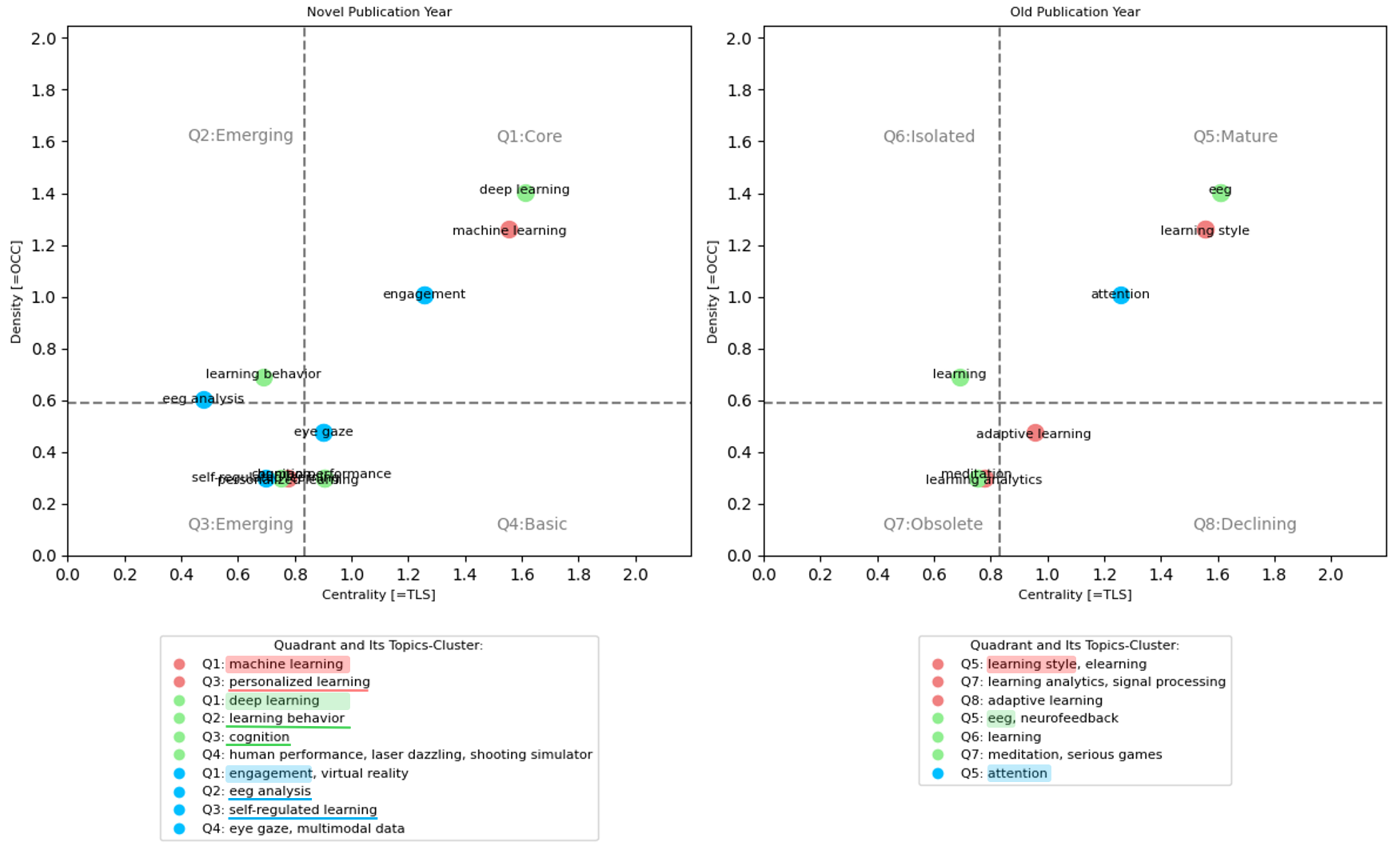}
  \caption{Topic of interest for RQ3 based on Enhanced Strategic Diagram \parencite{shafin2022thematic}}
  \label{fig:diagram}
\end{figure}

The second group of subjects—personalized learning, learning behavior, cognition, EEG analysis, and self-regulated learning—has led to the formulation of RQ3.2: What unexplored research gaps and areas offer potential for future EEG applications in education? Currently, the primary focus of the research is the need for systems that can adapt to learners' cognitive states and provide neurofeedback-based interventions to enhance self-regulation \parencite{dursoEnhancingEducationalOutcomes2024b,hanMultimodalLearningAnalytics2023,mohamedFacilitatingClassroomOrchestration2020}. Besides, we need to connect the brain signals with the behaviors of the learners, to extend the cognitive process measurement further than just the attention and working memory, and to perfect the reliability and generalizability of the predictive models so they can work with different kinds of learners \parencite{waznyRealTimeCognitiveStateNeuroimaging2018,zhangDesignImplementationEEGBased2021h,waznyRealTimeCognitiveStateNeuroimaging2018}. These voids reveal the potential for matching EEG technical progress in analyzing brain activity with teachers' objectives, thereby enabling the development of more personalized, adaptive, and learner-centered educational environments.\\

The quadrant-based analysis of RQ3 gives us a picture of the changes not only in the subject matter of educational research based on EEG but also in the same research's method. The presence of core and mature themes is evidence of a trend toward solid computational and neuroscientific approaches for engagement, attention, and learning style, whereas emerging themes suggest that there are areas that are not sufficiently explored, such as personalized learning, cognition, and self-regulation. Together, these revelations provide a guide for the further development of the application of EEG in education and making use of neural measures as a means for achieving the optimization of learning outcomes.\\

\section{Discussion}
\subsection{Methodological Approaches in EEG-Based Educational Research (RQ1)}
\subsubsection{Computational Techniques (RQ1.1)}
RQ1.1: What computational techniques are employed in EEG-based educational research, and how have they evolved? To address this question, we must investigate the trend and the present use of various computational techniques such as machine learning (ML), deep learning (DL), and classification. Principal to converting EEG readings into understanding of learning, cognitive conditions, and student participation are these three techniques \parencite{chenApplicationElectroencephalographySensors2024,elgammalNeuroTutorNeuralDecoding2025,yuvarajMachineLearningFramework2024a}. An overview of their evolution gives an idea of the underlying methodological trends and the selection of methods in educational research using EEG.\\

Initially, educational EEG research was based on using traditional machine learning algorithms like SVM, RF, KNN, Naive Bayes, and MLP to perform classification tasks \parencite{chowdhuriRealtimeClassificationEEG2024,dursoEnhancingEducationalOutcomes2024b,elkerdawyBuildingCognitiveProfiles2020,yuvarajMachineLearningFramework2024a}. These methods were used to classify attention, learning preferences, and cognitive conditions and generally involved a lot of preprocessing and manual feature extraction \parencite{chenApplicationElectroencephalographySensors2024,jawedDeepLearningBasedAssessment2024b}. Furthermore, ensemble methods such as bagging were also employed to increase the accuracy of the prediction \parencite{chowdhuriRealtimeClassificationEEG2024}, thus illustrating the role of classical ML as a basis for EEG study.\\

Convolutional neural networks (CNNs) influence this architecture, and there are also recurrent neural networks (RNNs), LSTM networks, LSTM-CNNs, and transformer-based methods that influence it. Classification accuracy can be improved through feature extraction using a deep learning algorithm approach. \parencite{chenApplicationElectroencephalographySensors2024, elgammalNeuroTutorNeuralDecoding2025,jawedDeepLearningBasedAssessment2024b,miahODLBCIOptimalDeep2024, yuvarajMachineLearningFramework2024a}. The incorporation of Bayesian optimization for hyperparameter tuning further led to a better performance of these models \parencite{miahODLBCIOptimalDeep2024}. On the other hand, the point has to be made that while DL is a mighty tool for automated feature engineering, some reports still imply that traditional ML can yield better results than DL if feature sets are well-developed \parencite{elkerdawyBuildingCognitiveProfiles2020}.\\

These results indicate the significant changes in the kind of computational techniques that have been used in educational research based on EEG. The area has moved from using the standard ML classifiers to using sophisticated DL and hybrid architectures, which not only shows the advancement of the methodology but also indicates the increasing demand for automated and high-performance classification of the cognitive and learning domains. Such an evolution is a clear indication that computational means are still finding a way to make the decoding of EEG signals more feasible and to provide the user with more practical and effective educational interventions \parencite{chenApplicationElectroencephalographySensors2024,elgammalNeuroTutorNeuralDecoding2025,jawedDeepLearningBasedAssessment2024b}.\\

\subsubsection{EEG Signal Processing and Preparation (RQ1.2)}
Knowledge of how EEG signals are handled and set up is a must-have skill to be able to peek into learners’ cognitive states, which is the basis of educational analysis. RQ1.2- How is the EEG signal processed and prepared to support meaningful educational analysis? Describes the procedural details of signal processing, EEG analysis, feature extraction, and feature selection, showing how these activities coalesce to meet the requirements of quality and relevance for further computational analysis \parencite{liuLearnerCognitiveState2024,mohamedFacilitatingClassroomOrchestration2020,yuvarajMachineLearningFramework2024a}.\\

The processing of the EEG signal normally commences with the data acquisition and preprocessing of the data to eliminate the addition of noise and artifacts \parencite{liuLearnerCognitiveState2024,sulaimanDevelopmentEEGBasedSystem2022}. Raw data is recorded using a variety of instruments, for instance, commercially available headphones such as the Emotiv EPOC or Emotiv Insight \parencite{arnaldoComputerizedBrainInterfaces2018,mohamedFacilitatingClassroomOrchestration2020}, and sophisticated 64-channel systems \parencite{liuLearnerCognitiveState2024}. Preprocessing entails filtering methods, for instance, bandpass, notch, and low-pass filtering methods to eliminate DC components, high-frequency noise, and powerline noise \parencite{chowdhuriRealtimeClassificationEEG2024,elgammalNeuroTutorNeuralDecoding2025,jawedDeepLearningBasedAssessment2024b,yuvarajMachineLearningFramework2024a}, and Independent Component Analysis (ICA), surrogate models, or automated reject are employed to eliminate artifacts introduced due to blinks of eyes, muscle activities, and cardiac activities \parencite{jawedDeepLearningBasedAssessment2024b,liuLearnerCognitiveState2024,yuvarajMachineLearningFramework2024a}.\\

To delineate the cognitive states, EEG preprocessing and feature extraction play a major role \parencite{arnaldoComputerizedBrainInterfaces2018,liBiosensorTechnologyAdaptive2025a,mohamedFacilitatingClassroomOrchestration2020}. In most cases, time-domain signals are changed to the frequency domain through Fast Fourier Transform (FFT) to get the Power Spectral Density (PSD) \parencite{chowdhuriRealtimeClassificationEEG2024,liBiosensorTechnologyAdaptive2025a,sulaimanDevelopmentEEGBasedSystem2022} and the five standard frequency bands—alpha, beta, theta, delta, gamma—are the ones that are studied to draw a relationship with cognitive states of the brain like alertness, relaxation, and concentration \parencite{chowdhuriRealtimeClassificationEEG2024,dursoEnhancingEducationalOutcomes2024b,sulaimanDevelopmentEEGBasedSystem2022}. Besides, coherence between different brain regions is also looked at to understand the extent of the interaction between the regions \parencite{dursoEnhancingEducationalOutcomes2024b}.\\

Among the feature extraction methods are statistics measures (mean, standard deviation, skewness, and kurtosis), Hjorth parameters, fractal dimensions, higher-order spectral features, Power Spectral Entropy (PSE), and Discrete Wavelet Transform (DWT) coefficients \parencite{elgammalNeuroTutorNeuralDecoding2025,jawedDeepLearningBasedAssessment2024b,liuLearnerCognitiveState2024,mohamedFacilitatingClassroomOrchestration2020,yuvarajMachineLearningFramework2024a}. Features of the brain, such as Concentration and Commitment, are calculated from power ratios of frequency bands, commonly utilizing a personalized baseline for each learner \parencite{dursoEnhancingEducationalOutcomes2024b}.\\

In the end, feature selection is the one that cuts down on the dimensions and picks out the most relevant features for the cognitive states modeling. The methods are Principal Component Analysis (PCA), Analysis of Variance (ANOVA), Sequence Forward Selection (SFS), and the combination of Genetic Algorithms with Random Forests for optimization \parencite{elgammalNeuroTutorNeuralDecoding2025,jawedDeepLearningBasedAssessment2024b,liuLearnerCognitiveState2024,yuvarajMachineLearningFramework2024a}. In certain papers, different feature sets are first tested for their effectiveness, and then the sets with the most discriminatory features are chosen for further analysis \parencite{elgammalNeuroTutorNeuralDecoding2025}. This multi-stage process—preprocessing, analysis, feature extraction, and selection—allows the creation of the predictive models that can both evaluate and forecast the cognitive states of the learners, which is a good example of the methodological rigor in EEG-based educational research \parencite{dursoEnhancingEducationalOutcomes2024b}.\\

EEG-based educational research, as a rule, aims at a broad variety of cognitive and affective states with the purpose of getting the whole picture of learners' attention, cognitive processes, and participation \parencite{cheahaEEGNeuralSubstrates2024,davidescoDetectingFluctuationsStudent2023,mohamedFacilitatingClassroomOrchestration2020}. The power of attention is emphasized in research, where sustained, selective, divided, and alternating attention, along with vigilance and attentional lapses, have been discussed by different learning styles \parencite{binabdulrashidDeterminationVigilanceboundLearning2012,chiangEEGBasedDetectionModel2018,christoforouYourBrainSTEM2024,davidescoDetectingFluctuationsStudent2023,elkerdawyBuildingCognitiveProfiles2020,grammerEffectsContextNeural2021a,niEEGBasedAttentionAnalysis2020a}. The researchers mention load as one of the cognitive states of the brain, besides working memory, perception, thinking, reasoning, and problem-solving, and they enable the reader to understand that the investigation of them is to measure the mental effort required during learning \parencite{cheahaEEGNeuralSubstrates2024,christoforouYourBrainSTEM2024,davidescoDetectingFluctuationsStudent2023,duAnalyzingEffectsInstructional2024b,elkerdawyBuildingCognitiveProfiles2020,goldbergEfficacyMeasuringEngagement2012,mohamedFacilitatingClassroomOrchestration2020}. Commitment to one thing or another, involving both a person's intellect and feelings, is studied through the prism of overall engagement, the latter being broken down into cognitive and emotional engagement, and further differentiated by related constructs such as focus, involvement, and flow \parencite{apicellaEEGbasedMeasurementSystem2022,christoforouYourBrainSTEM2024,dursoEnhancingEducationalOutcomes2024b,goldbergEfficacyMeasuringEngagement2012,natalizioRealtimeEstimationEEGbased2024,sandhuEvaluationLearningPerformance2017}.\\
\subsection{Educational Constructs and Learning-Related Phenomena in EEG-Based Educational Research (RQ2)}
\subsubsection{Cognitive and Affective States (RQ2.1)}

Investigating RQ2.1, the most typical cognitive and affective states, as the main question, the paper describing EEG usage in the educational field, brain states monitoring through EEG, attentional processes by EEG, cognitive states investigation by EEG, engagement measurement using EEG, affective responses detection through EEG, and behavior recording by EEG techniques is discussed here. These five psychosocial constructs open a window to the educational phenomena most frequently uncovered by EEG studies and can also be a valuable source to foresee the use of EEG-derived data in the design of personalized and adaptive learning interventions \parencite{cheahaEEGNeuralSubstrates2024,davidescoDetectingFluctuationsStudent2023,mohamedFacilitatingClassroomOrchestration2020}.\\

Affective states, such as emotions, motivation, and stress, are the focus here, and the authors analyzed these states together with arousal, delight, frustration, calmness, and tiredness \parencite{apicellaEEGbasedMeasurementSystem2022,deenadayalanEEGBasedLearners2018a,dursoEnhancingEducationalOutcomes2024b,goldbergEfficacyMeasuringEngagement2012,sandhuEvaluationLearningPerformance2017}. Moreover, the investigation of learning behavior is associated with EEG patterns and instructional strategies. Learning behavior describes the learning processes, which can be the rate of learning, the choice of the preferred learning style, and the transfer of skills \parencite{binabdulrashidDeterminationVigilanceboundLearning2012,chiangEEGBasedDetectionModel2018,christoforouYourBrainSTEM2024,davidescoDetectingFluctuationsStudent2023,deenadayalanEEGBasedLearners2018a,niEEGBasedAttentionAnalysis2020a,waznyRealTimeCognitiveStateNeuroimaging2018}. The researchers' motive is to grasp how different learner conditions, and to what extent, can affect educational outcomes with the help of such cognitive, affective, and behavioral insights, as well as to help in the pioneering of the adaptive learning environment \parencite{elkerdawyBuildingCognitiveProfiles2020,goldbergEfficacyMeasuringEngagement2012,mohamedFacilitatingClassroomOrchestration2020}.\\

These studies largely show that EEG offers a strong method to explore different cognitive and emotional states and their impact on learning processes. The understanding gained from the observation of attention, cognitive load, engagement, emotions, motivation, and learning behavior leads to the development of smart educational environments customized to the users’ requirements \parencite{elkerdawyBuildingCognitiveProfiles2020,goldbergEfficacyMeasuringEngagement2012,mohamedFacilitatingClassroomOrchestration2020}.

\subsubsection{Individual Differences and Learning Outcomes (RQ2.2)}
To answer RQ2.2—How is EEG utilized to assess individual differences and learning outcomes? —This discussion features the application of EEG to the detection of learning styles, to the assessment of cognitive states, and to the facilitation of personalized learning. When complemented by behavioral data, the EEG signals provide researchers and educators a venue to "look" into brain processes; thus, they can move further from the self-reporting method of individual learning differences \parencite{alhasanExperimentalStudyLearning2018,arnaldoComputerizedBrainInterfaces2018,cordovaIdentifyingProblemSolving2015,zhangDesignImplementationEEGBased2021h}.\\

EEG has been successfully utilized in the identification of learning styles as well as in capturing individual differences in cognitive preference \parencite{abdulrashidClassificationLearningStyle2010b,binabdulrashidSummativeEEGAlpha2014a,hasibuanProposedModelDetecting2025b,jawedDeepLearningBasedAssessment2024b,megataliLearningStyleClassification2014d,zhangDesignImplementationEEGBased2021h}. The changes in EEG can be very helpful to characterize one's visual and f.e. aural learners, thus machine learning classifiers like KNN and Linear Regression, which work with EEG data, were found to achieve very good performance in style detection \parencite{hasibuanProposedModelDetecting2025b,jawedDeepLearningBasedAssessment2024b,megataliLearningStyleClassification2014d,zhangDesignImplementationEEGBased2021h}. Cluster analysis of resting-state alpha and beta bands is also valuable in the identification of individual learning patterns, as it can uncover subtle differences in the understanding of learner differences \parencite{abdulrashidClassificationLearningStyle2010b,binabdulrashidSummativeEEGAlpha2014a}.\\

The use of EEG has been extended to learning evaluation in which attention, engagement, and cognitive workload are monitored as these factors are closely related to academic outcomes \parencite{arnaldoComputerizedBrainInterfaces2018,chiangEEGBasedDetectionModel2018,dursoEnhancingEducationalOutcomes2024b,niEEGBasedAttentionAnalysis2020a}. The changes in brainwaves, including the changes in theta and alpha bands, give a picture of one’s mental effort and the recruitment of resources during a learning task, while other parameters such as focus, involvement, relaxation, fatigue, and stress contribute to the identification of the best cognitive states, like the "Flow state" \parencite{alhasanExperimentalStudyLearning2018,deenadayalanEEGBasedLearners2018a,dursoEnhancingEducationalOutcomes2024b}. These EEG-based signals provide the possibility for teachers to revise the way they deliver their content, change the level of difficulty of the content on the fly and even create the type of learning environments that would be compatible with the cognitive capacities and preferences of the individual learners \parencite{arnaldoComputerizedBrainInterfaces2018,binabdulrashidSummativeEEGAlpha2014a,cordovaIdentifyingProblemSolving2015,zhangDesignImplementationEEGBased2021h}.\\

EEG has been a great tool in identifying individual variances and learning results that eventually create a robust basis for personalized learning. The combination of data on learning styles, brain states, and participation levels aids the creation of flexible learning settings that can improve the learning journey and the effectiveness of the different types of learners \parencite{alhasanExperimentalStudyLearning2018,arnaldoComputerizedBrainInterfaces2018,cordovaIdentifyingProblemSolving2015,dursoEnhancingEducationalOutcomes2024b,zhangDesignImplementationEEGBased2021h}.\\

\subsubsection{Emerging Educational Environments and Interventions (RQ2.3)}
EEG implementation in modern educational settings and changes has become popular, notably through virtual reality (VR) and neurofeedback. These methods use one’s brain activity data to customize the learning experience, improve one’s focus and participation, and promote the development of one’s own control. Through connecting brain function checking with adjustable learning programs, EEG is a strong connector between brain functions and new teaching methods \parencite{hubbardEnhancingLearningVirtual2017,murdochExperientialLearningBasedApproach2019,waznyRealTimeCognitiveStateNeuroimaging2018}.\\

One of the main uses is EEG monitoring of a person's brain activity and VR environment adaptation accordingly. The VR system, by detecting changes in the focus or involvement of the user, can be altered during use, thus creating a closed-loop feedback system that gives the user the possibility to maintain his/her concentration and to optimize the learning process \parencite{hubbardEnhancingLearningVirtual2017,limLearningAttentionImprovement2018}. Moreover, EEG also allows the integration of neurofeedback, which is a tool that gives instant information about the cognitive states of the users. In these game-based or simulation-driven activities, students can drill attention and self-control. They are, thus, making the brain functions, which are usually considered abstract and difficult to grasp, into tangible and interactive ones \parencite{handayaniConceptPaperAlert2024,krellSchoolBasedNeurofeedbackTraining2023,niEEGBasedAttentionAnalysis2020a}.\\

Adaptive learning can be greatly expanded with combined EEG and VR. The changes could be character appearances, the level of challenge, or even the environment in the simulation, all depending on the user's brain activity. These changes result in users feeling more involved and having a unique experience \parencite{torres-rodriguezGamifiedSimulatorsEEG2025,tremmelEstimatingCognitiveWorkload2019a}. Moreover, the use of portable EEG devices is steadily growing in the field of online learning and this trend points towards a future where the concentration level of the learners can be gauged in real-time and they can be given feedback both visually and through signals so that their engagement level is maintained \parencite{elgammalNeuroTutorNeuralDecoding2025,hanMultimodalLearningAnalytics2023,wangEvaluateLearningAttention2022}. These changes enable the tracking of mental states objectively while the adaptive interventions are provided, which is usually a challenge when using traditional behavioral measures.\\

By using EEG together with VR and neurofeedback, educational research is changing. It is no longer only observing but designing the environments that are adaptive, responsive, and learner centered. Such a combination not only facilitates a deeper understanding of attention, engagement, and cognitive load but also assists in the creation of systems that can interact directly with the brain signals. As the developments in the field of EEG continue, this technology is becoming a very important connection between educational and neuroscience innovations, thus providing new ways of matching learning with the cognitive abilities of each person.\\

\subsection{Emerging Trends, Research Gaps, and Future Directions in EEG-Based Educational Research (RQ3)}
\subsubsection{Thematic and Methodological Trends (RQ3.1)}
One of the significant shifts in the use of EEG to study the educational processes is the changed scope of the research itself and the way the scientists treat cognitions in the same learning environment. The study of cognitive processes in the learning environment has changed significantly, including the thematic and methodological shifts.\\

In dealing with RQ3.1, the recently published journal focuses on engagement, attention, and learning styles as being crucial, while methodological changes—such as from wearable EEG devices to algorithmic learning and deep learning—have also extended the effects and the range of results to be found. Such changes demonstrate how the field has moved from largely exploratory studies to the use of more mature, data-centric approaches for the adaptation and personalization of educational programs.\\

Conceptually, the identification of engagement and attention as core phenomena has been the result of their being mentioned most frequently among the examined constructs in the studies in question, and also, their being the main factors that affect the learning outcomes. These studies have highlighted the role of attentiveness, concentration, and task-related focus as three of the most important indicators when it comes to adapting instruction and achieving the best classroom practices \parencite{chowdhuriRealtimeClassificationEEG2024,elkerdawyBuildingCognitiveProfiles2020,sandhuEvaluationLearningPerformance2017}. Correspondingly, the study of learning styles has also developed substantially, based on the different instructional strategies that lead to changes in the brain activity for attention and engagement in the real-world classroom setting \parencite{grammerEffectsContextNeural2021a,niEEGBasedAttentionAnalysis2020a}.\\

These advances in themes have been made possible through methodological innovations. The use of transportable EEG systems, for example, the Emotiv EPOC, allows for the acquisition of data in naturalistic settings so that teachers can capture the brain without interrupting classroom processes \parencite{goldbergEfficacyMeasuringEngagement2012,grammerEffectsContextNeural2021a}. EEG frequency band analyses of alpha, beta, and theta serve as indicators of changes in cognitive states and levels of engagement \parencite{cheahaEEGNeuralSubstrates2024}. In the computational level, machine learning has come to the forefront of high-dimensional EEG data processing with the help of algorithms such as SVM, Random Forests, and Naïve Bayes for efficient classification of attention and engagement states \parencite{dursoEnhancingEducationalOutcomes2024b,sandhuEvaluationLearningPerformance2017}. Recently, CNNs and LSTMs have come to further improve the field of study through automatic feature extractions and EEG signal modeling of temporally dependent variables \parencite{apicellaEEGbasedMeasurementSystem2022,chenApplicationElectroencephalographySensors2024}.\\

The combination of these methods, multimodal data fusion, is an important method pattern to combine EEG with eye-tracking, facial recognition, or video to get a more complete understanding of the cognitive and emotional states \parencite{christoforouYourBrainSTEM2024,zhangMultimodalFastSlow2023}. Such a merge is instrumental for the implementation of the likes of adaptive learning systems, brain–computer interfaces, and neurofeedback-driven interventions, where the on-the-spot observation guides individualized feedback and learning adaptation \parencite{handayaniConceptPaperAlert2024,natalizioRealtimeEstimationEEGbased2024}.\\

The combination of thematic emphases and methodological innovations has largely influenced the journey of educational research based on EEG, which is now situated at the crossroads of neuroscience, computing, and pedagogy. The scholars can be said to be building the infrastructure for learning environments that are more adaptive and individualized by integrating insights into engagement, attention, and learning styles with the use of advanced analytical methods such as deep learning and multimodal fusion. This development marks the pioneering end of the field's ongoing transformation and offers the basis for further research directions to be explored later.\\

\subsubsection{Research Gaps and Underexplored Areas (RQ3.2)}
Though the EEG-based research in the field of education has increased dramatically, there are still some neglected areas that could considerably expand the domain. These areas relate to personalization of education, the students' learning habits, their cognitive processes, the analysis of EEG, and self-regulated learning. Detecting such gaps and addressing them, considering the proper research, is the key to the use of EEG as a tool for facilitating flexible, data-driven learning environments.\\

It is important to note that present studies in the field of individualized education have only delved into the aspects of real-time adjustment of instruction based on learners' mental states \parencite{dursoEnhancingEducationalOutcomes2024b,mohamedFacilitatingClassroomOrchestration2020}. The main emphasis of future research should be on the development of adaptive learning systems that can alter the content on the spot as per the present attention span or working memory, and also on the creation of neurofeedback devices that cater to the needs of a single person \parencite{hanMultimodalLearningAnalytics2023}. The identification of unique learners' personality traits and the use of EEG-based systems in teaching different subjects would facilitate the transfer of these intervention methods to new areas \parencite{liBiosensorTechnologyAdaptive2025a}.\\

Research on learning behavior has major opportunities for the future. Though EEG is a successful tool for recording the changes in attention, the connection between these neural signals and the behavioral patterns that can be observed is still quite significantly unclear \Parencite{mohamedFacilitatingClassroomOrchestration2020}. The research should find out the relationship between the brain signals that indicate continuous attention and the characteristics of human behavior, such as the capacity to persist with a task, the degree to which a person is involved in a particular activity, and the use of learning strategies especially when the learners are dealing with different kinds of multimedia \parencite{hanMultimodalLearningAnalytics2023}. More emphasis on the development differences between various age groups, including the differences between the younger and the older learners, is also necessary \parencite{waznyRealTimeCognitiveStateNeuroimaging2018}.\\

Most research in the area of cognition has been very focused on the processes of attention and working memory, with a lot of other critical cognitive processes being neglected. The research should use EEG to monitor cognitive abilities like perception, coordination, and reasoning \parencite{mohamedFacilitatingClassroomOrchestration2020}. The improvement of the EEG-based indicators to separate the good cognitive challenge from the bad stress and to make stronger connections between the EEG signals and long-term learning outcome is also an important issue \parencite{waznyRealTimeCognitiveStateNeuroimaging2018}. Studying different neural sources of various learning styles can open up new ways to understand differences in human cognition \parencite{zhangDesignImplementationEEGBased2021h}.\\

Methodological advancements in EEG analysis are a must-have to solve the technical and interpretative challenges. Noise and motion artifacts continue to be the main barriers \parencite{waznyRealTimeCognitiveStateNeuroimaging2018}; thus, more advanced preprocessing methods and stronger filtering algorithms are needed \parencite{liBiosensorTechnologyAdaptive2025a}. The use of deep learning on raw EEG data combined with multimodal biometric integration can open up new paths for the recognition to be more accurate \parencite{mohamedFacilitatingClassroomOrchestration2020,zhangDesignImplementationEEGBased2021h}. It is still a very important next step to work on making predictive models that are dependable and can be generalized to different populations \parencite{dursoEnhancingEducationalOutcomes2024b}.\\

Self-regulated learning is an important aspect that, however, is still largely unexploited. For instance, learners may be enabled to track their brain activities with real-time EEG-based feedback and hence gauge their cognitive states, which could motivate them to adjust their learning techniques in a more efficacious way \parencite{mohamedFacilitatingClassroomOrchestration2020}. Moreover, extending neurofeedback as a feasible instrument in diverse educational settings, such as portable and user-friendly applications, may break down barriers to its use and facilitate an increase in learners' freedom \parencite{chiangEEGBasedDetectionModel2018,hanMultimodalLearningAnalytics2023}.\\

Considering these gaps, they signal the need for a type of research that not only focuses on technical EEG advancements but also on educational objectives, which include personalization, the cognitive domain, and learner agency. Therefore, by going deeper into these wider domains than only focusing on attention, future studies can widen the scope and achieve more profound educational transformations. Overcoming these issues will be very important in achieving the promise that EEG is a revolutionary instrument in the creation of education that is adaptive and centered on the learner.

\subsection{Contribution, Implication, and Limitation}
This research study adds to EEG-based instructional research with the help of procedural and applied insights that broaden the understanding of learners' cognitive and emotional states. What is more, it presents a brand-new technique, BenSLR, which is more of an innovative character as it merges Bibliometric Analysis (BA) directly within the Systematic Literature Review (SLR) context than those studies where BA and SLR were dealt with separately. The main idea behind this approach is to combine various advanced computational techniques, EEG signal processing, and neurofeedback insights, which in turn deepen the researchers' understanding of attention, engagement, and personalized learning. The element of individual differences, such as learning style tendencies, in this architecture serves as a supporting feature for the general aim of adjusting the instruction to the needs of learners rather than being a separate focus. This holistic view further facilitates the identification of EEG-based educational research trends, methods, and findings.\\

The implications of these findings span both practice and research. EEG-guided systems, alongside BenSLR insights, are able to steer the development of adaptive learning environments whose content, difficulty of tasks, and feedback are adaptively altered on an ongoing basis in alignment with learners' cognitive states. Another intervention of VR and neurofeedback also allows learners a real-time capacity to self-regulate focus, attention, and engagement. Incorporating differences on an individualized basis, including learning style patterns, enables these adaptive approaches within a holistic framework of personalized and student-centric education.\\

Despite the limitations, this research has boundaries. It is solely dependent on publications indexed by Scopus. This might exclude some of the relevant investigations and introduce some bias. Differences in EEG devices, preprocessing, and analysis protocols, for example, across different works might result in a limitation of the generalizability of the findings. Researchers may consider integrating various databases in future work to improve the reliability and reach of the findings by using multimodal EEG data and cross-contextual validation. Nevertheless, the work lays out a solid conceptual framework for the implementation of attention-guided, adaptive, and personalized learning strategies, thus providing the methodological aspects as an insight for educational research based on the EEG.\\

\section{Conclusion}
The research effort combined EEG-based educational neuroscience studies to clarify parameters of the educational sector, the core concepts put forward, and the change in the learning process. Their strategy involved identifying, selecting, and analyzing relevant studies to answer specific research questions. The results demonstrate that engagement, attention, and learning styles continue to be significant concepts that researchers utilize when applying machine learning and deep learning to predict the cognitive state. Besides that, EEG signal processing technical methods, cognitive and affective measures, along with new potential treatments, for example, virtual reality and neurofeedback, have been acknowledged as the leading trends that mirror changes in the methodology and theme of the area.\\

The definition of cognitive processes within attention and working memory is the starting point for present research to build adaptive learning systems that can adjust to the learners' cognitive states and learning styles, which are disclosed during the learning process in real-time. The reliability and generalizability of the predictive models across different populations of learners are among the suggested directions for future research, these directions presenting ways for the implementation of EEG indications to facilitate more personalized education, which is the central point of the learner. By illustrating the applicability of the BenSLR method, this research also constitutes a methodological framework for future systematic reviews, thus facilitating evidence-based decision-making and the planning of novel EEG-based educational-sector interventions and endowing a flexible approach that is compatible with other issues and research areas.\\

\section*{Acknowledgement}
The Ministry of Higher Education, Science, and Technology of the Republic of Indonesia has contributed to this study through the Regular Fundamental Research grant scheme under the grant number 0419/C3/DT.05.00/2025. The Indonesia Endowment Fund for Education (LPDP) also provided some assistance. The authors are thrilled to express their sincere thanks to those who have helped and supported this study.\\

\section*{Author Contribution}
AW was instrumental in the conceptualization, methodology, and formal analysis, as well as in the writing of the original draft and in the review and editing. SH's role was mainly concerned with the validation and review and editing of the manuscript. WMB was involved in the formal analysis and the writing of the original draft. RD oversaw the investigation and visualization. RP took care of data curation and managed resources. CS was involved in writing, reviewing, and editing. The final manuscript has been read and approved by all authors for publication.

\printbibliography

\end{document}